Derivation of CPT resonance signals from density-matrix equations with all relevant sublevels of Cs atoms and confirmation of experimental results

Kenta Matsumoto[1,2], Sota Kagami[1,2], Takahiro Fujisaku[1,2], and Akihiro Kirihara[1,2]

[1]Secure System Platform Research Laboratories, NEC Corporation, 1753 Shimonumabe, Nakahara-ku, Kawasaki, Kanagawa 211-0011, Japan

[2]National Institute of Advanced Industrial Science and Technology (AIST), NEC-AIST Quantum Technology Cooperative Research Laboratory, 1-1-1 Umezono, Tsukuba, Ibaraki 305-8568, Japan

Shinya Yanagimachi[3]

[3]National Institute of Advanced Industrial Science and Technology (AIST), 1-1-1 Umezono, Tsukuba, Ibaraki 305-8563, Japan

Takeshi Ikegami[4] and Atsuo Morinaga[4]

[4]Micromachine Center, AIST Tsukuba East 4G, 1-2-1 Namiki, Tsukuba, Ibaraki 305-8564, Japan

(Abstract)

Coherent-population-trapping resonance in vapor cell is a quantum interference effect that appears in the two-photon transitions between the ground-state hyperfine levels of alkali atoms and is often utilized in miniature and centimeter-scale clock devices. To quantitatively understand and predict the performance of this phenomenon, it is necessary to consider the transitions and relaxations between all hyperfine Zeeman sublevels involved in the different excitation processes of the atom. In this study, we constructed a computational multi-level atomic model of the Liouville density-matrix equation for 32 Zeeman sublevels involved in the $D_1$ line of $^{133}Cs$ irradiated by two frequencies with circularly polarized components and then simulated the amplitude and shape of the resonance spectrum of the transmitted light through centimeter-scale Cs vapor cells. We show that the numerical solutions of the equation and analytical investigations adequately explain a variety of the characteristics observed in the experiment.

I. INTRODUCTION

Coherent-population-trapping (CPT) resonance is a quantum interference phenomenon observed using a two-photon Λ-type transition between the ground hyperfine Zeeman sublevels of an alkali atom [1,2]. Thanks to its high Q factor at the microwave transition, it is currently utilized as a key spectroscopic technique for creating portable atomic clock devices [3]. In more than 20 years following the first detection of CPT resonance in vapor cells [4,5] and proposal for microfabricated atom vapor cells [6-8], ongoing advancements have led to the development of miniature atomic clocks using silicon micromachining and semiconductor laser technology [9], the achievement of high contrast signal [10], and increased robustness of the resonance frequency to fluctuations in the external environment and the excitation light itself [11,12].

The optical excitation scheme originally utilized in the CPT clock is a circular polarization for bichromatic excitation lights, $\sigma^- - \sigma^-$ or $\sigma^+ - \sigma^+$, but such polarization pumps a significant fraction of the atoms into Zeeman edge (trap) states, thus reducing the contrast of the CPT signal. To prevent this reduction, excitation schemes such as push-pull optical pumping [13], counter-propagating $\sigma^+ - \sigma^-$ polarization [14], a pair of orthogonal linear polarizations (Lin $\perp$ Lin) [15], or a pair of parallel linear polarizations (Lin $\parallel$ Lin) [16] have been proposed. The first three methods produce the CPT resonance between two hyperfine states $|F_g = 3, m = 0\rangle$ and $|F_e = 4, m = 0\rangle$ on each leg of the $\Lambda$ scheme, namely (0, 0) CPT resonance [17]. This resonance is essentially a double-$\Lambda$ scheme, in which a dark state common to the two $\Lambda$ schemes exists [18]. In contrast, no (0, 0) CPT resonance occurs in the fourth method because the dark state for one $\Lambda$ scheme is the bright state for the other. Instead, two CPT resonances are produced between $|F_g, m = -1\rangle$ and $|F_e, m = 1\rangle$ and between $|F_g, m = 1\rangle$ and $|F_e, m = -1\rangle$, namely (–1, 1) and (1, –1) CPT resonances. These doublet resonances split in frequency due to the second order Zeeman effect [19]. The relationship between the CPT resonance and excitation polarization scheme on $^{133}$Cs atoms has been studied by Liu et al. [17].

In our previous work [20], we showed that the amplitudes of the (–1, 1) and (1, –1) CPT resonances excited with Lin $\parallel$ Lin polarization increase approximately in proportion to the excitation intensity, while in contrast, the amplitude of the (0, 0) CPT resonance excited with $\sigma^- - \sigma^-$ polarization moderately saturates. We claimed that the former is best described by a simple three-level model and the latter by a four-level model with a trap state. Up to now, the symmetry, width, and frequency shift of the resonance spectrum related to the optical detuning of the excitation lights were investigated using few-level models [21, 22]. However, a model that includes all Zeeman sublevels will be effective for quantitatively understanding the amplitude, the width and the shape of the CPT resonances [23]. These behaviors of the CPT spectrum can be better understood by solving the Liouville density-matrix equation taking into account 32 Zeeman sublevels related to the $D_1$ line of $^{133}$Cs.

When constructing a model with 32 Zeeman sublevels, it is necessary to configure a detailed relaxation process between 16 ground levels, though it is simple in the three-level model because there are only two ground levels. In 2017, Warren et al. developed an atomic model using the Liouville density-matrix equation taking into account all relevant 16 Zeeman sublevels in the $D_1$ line of $^{87}$Rb atoms and compared the calculated results with the corresponding experimental results for excitation with three different polarization configurations [24]. They assumed a uniform relaxation process between the magnetic sublevels of the ground states. In an alternative approach, Matsuda et al. utilized the magnetic dipole relaxation between the magnetic sublevels of the ground states in the $D_1$ line of $^{133}$Cs atoms [25].

We make two key contributions in the current work. First, we construct a multi-level atomic model of the Liouville density-matrix equation [23-26] for studying CPT resonances formed by the bichromatic lights of various excitation schemes in the manifold of $^{133}$Cs atoms. Second, using the constructed model, we simulate the amplitude and shape of the CPT resonance excited by different polarizations, frequencies, and intensities of the excitation lights, and elucidate the underlying mechanisms by comparing them with the corresponding experimental results. In Sec. II.A, we derive the multi-level atomic model using the density matrix equation, and in II.B, we show the formulations for the line shape, width, and light shift of the CPT resonance spectrum guided from the present multi-level atomic model. Section III describes our experimental setup and Cs vapor cells with buffer gas. In Sec. IV, we compare the experimental results with the calculated results. IV.A reports how the Zeeman CPT spectra with different buffer gas pressures vary due to the relaxation process. IV.B shows that the amplitude of the first order Zeeman CPT spectra depends on the common detuning frequency of the excitation light. In IV.C, we clarify that the $(m, m)$ CPT resonance appears for Lin $\parallel$ Lin excitation (except for (0, 0)) in the second order Zeeman CPT spectrum and discuss the required conditions to prohibit the CPT resonances of double-$\Lambda$ schemes. IV. D explains how the amplitude of the CPT resonance for $\sigma^- - \sigma^-$ excitation saturates while that for Lin $\parallel$ Lin excitation increases in proportion to the excitation intensity depending on the variation of the population in the trap state. We conclude in Sec. V with a brief summary. Appendix provides additional detail how the shape of the CPT spectrum is rigorously derived from the 32-level model.

## II. FORMULATION AND CALCULATION

### A. Liouville equations for CPT spectrum

We aim to construct a multi-level atomic model including all the Zeeman sublevels in the $D_1$ transition of $^{133}$Cs so as to theoretically investigate the CPT resonances excited with various polarizations. Figure 1 shows the energy structure of the hyperfine Zeeman sublevels in the $D_1$ line of $^{133}$Cs under a magnetic field as a perturbation together with the definition of the energy detuning of the bichromatic excitation lights for the CPT resonance, whose angular frequencies are $\omega_1$ and $\omega_2$. In this paper, all quantities of the energy detuning and energy levels are given in the unit of angular frequency. We designate sublevels belonging to $6S_{1/2}$ $F = 3$, $F = 4$, and $6P_{1/2}$ $F' = 3$ or $F' = 4$ as $g$, $e$, and $i$, respectively. The unperturbed energies of the hyperfine ground states, $6S_{1/2}$ $F = 3$ and $F = 4$, are represented as $\omega_g^0$ and $\omega_e^0$, respectively. $\omega_{i0}$ is the mean energy between the unperturbed energies of $6P_{1/2}$ $F' = 3$ and $F' = 4$. We define the propagation direction of the bichromatic lights is z-axis and the quantized magnetic field with strength $B$ is applied along the direction of the light propagation. We also define energies of the Zeeman sublevels under the magnetic field $B$ as $\omega_g^B$, $\omega_e^B$ and $\omega_i^B$, which are shifted by the Zeeman effect according to Breit-Rabi's formula [19]. Then, there are 32 Zeeman sublevels in the hyperfine ground states, $6S_{1/2}$ $F = 3$ and $F = 4$, and in the hyperfine excited states, $6P_{1/2}$ $F' = 3$ and $F' = 4$, some of which are coupled by the $D_1$ transition depending on the polarization of excitation lights. As shown in Fig. 1, Zeeman sublevels are designated from 1 to 32 in order of the magnetic quantum numbers $m_F$ from $-F$ to $F$, and from $g$ to $i$.

The two photons $\omega_1$ and $\omega_2$ $(< \omega_1)$ induce a three-level $\Lambda$-type CPT resonance between the two hyperfine levels in the ground state and an excited level. The energy detuning of $\omega_1$ from the related $D_1$ transition energy is $\Delta_1 = \omega_1 - (\omega_{i0} - \omega_g^0)$. Similarly, that of $\omega_2$ is $\Delta_2 = \omega_2 - (\omega_{i0} - \omega_e^0)$. The Raman detuning of $\omega_1 - \omega_2$ from the ground hyperfine splitting $(\Delta_{hfs} = \omega_e^0 - \omega_g^0)$ is given by $\Delta_R = (\omega_1 - \omega_2) - \Delta_{hfs}$. In the CPT resonance experiment, generally, the +first and −first sidebands generated from a single laser source by modulating at a frequency of $\left(\Delta_{hfs} + \Delta_R\right)/2$ are used as the two photons $\omega_1$ and $\omega_2$. As shown in Fig. 1, we define the common detuning $\Delta_{opt}$ as $\Delta_{opt} = \Delta_1 - \dfrac{\Delta_R}{2} = \Delta_2 + \dfrac{\Delta_R}{2}$.

The system with a set of the orthonormal quantum states $|1\rangle - |32\rangle$ corresponding to the 32 magnetic sublevels of a Cs atom (presented in Fig. 1) is described by the Heisenberg equation of the density matrix component $\rho$ and the interaction Hamiltonian $H$, as

$$\frac{\partial}{\partial t}\rho = \frac{1}{i\hbar}[H,\rho] \tag{1}$$

where $\hbar$ is the Planck constant divided by $2\pi$. The coupling of the atomic states to two coherent radiation fields is described within the rotating wave approximation, as

$$\frac{1}{\hbar}H = \sum_{l=1}^{32}\delta_l|l\rangle\langle l| - \frac{1}{2}\sum_{s=1}^{16}\sum_{t=17}^{32}\left(\Omega_{st}|s\rangle\langle t| + \Omega_{st}^*|t\rangle\langle s|\right) \tag{2}$$

where

$$\delta_l = \begin{cases} \omega_l^B - \omega_g^0 + \dfrac{\Delta_R}{2} & l = 1,\cdots,7 \\ \omega_l^B - \omega_e^0 - \dfrac{\Delta_R}{2} & l = 8,\cdots,16 \\ \omega_l^B - \omega_{i0} - \Delta_{opt} & l = 17,\cdots,32 \end{cases} \tag{3}$$

Here, the coupling term $\Omega_{st} = \langle s|(-\mathbf{d}\cdot\mathbf{E})/\hbar|t\rangle = \Omega_{ts}^*$ is the Rabi frequency, where $\mathbf{d}$ is the electric dipole moment and $\mathbf{E}$ is the electric field of the excitation lights.

First, as the excitation light we consider a circularly polarized $\sigma^+$ light with the amplitude of $E(t)$, which propagates along the z-axis. The Rabi frequency for $\sigma^+$ excitation light is defined as

$$\Omega_{st}^{\sigma^+} = -\frac{E(t)}{\hbar}d_{FF'}\langle F',m_F'|F,1,m_F,1\rangle. \tag{4}$$

Similarly, for a circularly polarized $\sigma^-$ light, the Rabi frequency is defined as

$$\Omega_{st}^{\sigma^-} = -\frac{E(t)}{\hbar}d_{FF'}\langle F',m_F'|F,1,m_F,-1\rangle. \tag{5}$$

Here, $F$ and $F'$ are the total angular momenta of $|s\rangle$ and $|t\rangle$, respectively. $m_F$ and $m_F'$ are the magnetic quantum numbers of $|s\rangle$ and $|t\rangle$, respectively. $d_{FF'}$ is the reduced matrix element of the dipole moment operator between levels whose total angular momenta are $F$ and $F'$, and $\langle F',m_F'|F,1,m_F,\pm 1\rangle$ is the Clebsch-Gordan coefficient. These values for the Cs atom are given in [27].

Next, as the excitation light $\mathbf{E}(t)$, we consider the linearly polarized light with the amplitude of $E(t)$ whose polarization forms an angle $\theta$ with the x-axis in the xy-plane. The linearly polarized electric field is rewritten by the superposition of two circular polarizations $\sigma^+$ and $\sigma^-$ using the spherical vector basis $\mathbf{e}_{\pm 1} = \mp(\mathbf{e}_x \pm i\mathbf{e}_y)/\sqrt{2}$, as

$$\mathbf{E}^{\text{lin}}(t,\theta) = \mathrm{E}(t)\left(\mathbf{e}_x \cos\theta + \mathbf{e}_y \sin\theta\right) = \frac{\mathrm{E}(t)}{\sqrt{2}}\left(-e^{-i\theta}\mathbf{e}_{+1} + e^{i\theta}\mathbf{e}_{-1}\right). \tag{6}$$

Using Eqs. (4) and (5), the Rabi frequency of the linear polarized light is rewritten as

$$\Omega_{st}^{\text{lin}} = -\frac{e^{-i\theta}}{\sqrt{2}}\Omega_{st}^{\sigma+} + \frac{e^{i\theta}}{\sqrt{2}}\Omega_{st}^{\sigma-}. \tag{7}$$

In the CPT resonance, the bichromatic excitation lights of $\omega_1$ and $\omega_2$ interact with two ground hyperfine states and one excited state by a $\Lambda$-type scheme. The bichromatic Rabi frequency in this case is also composed of two Rabi frequencies $\Omega_{1st}(\mathrm{E}_1(t))$ and $\Omega_{2st}(\mathrm{E}_2(t))$, where $\mathrm{E}_1(t)$ and $\mathrm{E}_2(t)$ are the amplitudes of the excitation lights of $\omega_1$ and $\omega_2$, respectively. Since the detuned frequency component disappears due to the rotating wave approximation, $\Omega_{2st}(\Omega_{1st})$ disappears for $s = 1,\cdots,7\,(8,\cdots,16)$.

Let us assume the direction of the electric field of $\omega_1$ is parallel to the x-axis, while that of $\omega_2$ forms an angle $\theta$ with the x-axis in the xy-plane. The coupling term is then written as

$$\Omega_{st}^{\text{lin}} = \begin{cases} \Omega_{1st}^{\text{lin}} = -\dfrac{1}{\sqrt{2}}\Omega_{1st}^{\sigma+} + \dfrac{1}{\sqrt{2}}\Omega_{1st}^{\sigma-} & s = 1,\cdots,7 \\[2mm] \Omega_{2st}^{\text{lin}} = -\dfrac{e^{-i\theta}}{\sqrt{2}}\Omega_{2st}^{\sigma+} + \dfrac{e^{i\theta}}{\sqrt{2}}\Omega_{2st}^{\sigma-} & s = 8,\cdots,16 \end{cases}. \tag{8}$$

In the Lin || Lin polarization scheme, we adopt $\Omega_{2st}^{\text{lin}}$ with $\theta = 0$. In the Lin $\perp$ Lin polarization scheme, which corresponds to the push-pull scheme, we adopt $\Omega_{2st}^{\text{lin}}$ with $\theta = \pi/2$.

The evolution of the atomic system is governed by the Liouville equation, namely, the equation of motion for the density operator $\hat{\rho} = \sum_{l=1}^{32}\sum_{m=1}^{32}\rho_{lm}|l\rangle\langle m|$, as follows.

$$\frac{\partial}{\partial t}\hat{\rho} = -\frac{i}{\hbar}\left(\hat{H}\hat{\rho} - \hat{\rho}\hat{H}\right) - \frac{i}{\hbar}\left(\hat{H}'_\Gamma\hat{\rho} + \hat{\rho}\hat{H}'_\Gamma\right) + \hat{\Lambda}, \tag{9}$$

where a non-Hermitian operator $\hat{H}'_\Gamma$ and a source matrix $\hat{\Lambda}$ are added to reflect the relaxation process. $\hat{H}'_\Gamma$ accounts for the decays of atomic states by defining

$$\frac{i}{\hbar}\hat{H}'_\Gamma = -\frac{i}{2}\sum_{l=1}^{32}\Gamma_l|l\rangle\langle l| = -\frac{i}{2}\left(\gamma_p\sum_{l=1}^{16}|l\rangle\langle l| + \Gamma\sum_{l=17}^{32}|l\rangle\langle l|\right), \tag{10}$$

where $\Gamma_l$ is the total decay rate of a sublevel $|l\rangle$ and is a value that depends on several experimental parameters (e.g., the pressure of buffer gas, temperature, and the coating situation of the cell wall). For simplicity, we assume that all excited states have the same decay rate $\Gamma$, namely, $\Gamma_l = \Gamma$ for $l = 17,\cdots,32$, which can be estimated from the profile of the absorption spectrum. Similarly, the relaxation rates of all ground hyperfine states are assumed to be $\gamma_p$, namely, $\Gamma_l = \gamma_p$ for $l = 1,\cdots,16$, which is estimated from the width of the CPT spectrum.

On the other hand, the source matrix $\hat{\Lambda}$ contains non-zero diagonal elements that account for the influx of atoms decaying from other states, as follows:

$$\hat{\Lambda} = \sum_{l=1}^{32}\Lambda_l\left(\Gamma_1\rho_{11},\Gamma_2\rho_{22},\cdots,\Gamma_{32}\rho_{3232}\right)|l\rangle\langle l|. \tag{11}$$

$\Lambda_l$ represents the total influx rate of $|l\rangle$, which is a function of the product of decay rate $\Gamma_m$ and the population

$\rho_{mm}$ of state $|m\rangle$. We ignore the influx rates into the excited states because these are by far smaller than those into the ground states; $\Lambda_l = 0$ for $l = 17, \cdots, 32$. We assume that the decay process from an excited state $|n\rangle$ to a ground state $|m\rangle$ is governed by cesium-nitrogen collisions, which would have the dipole-quadrupole interaction, whose decay rates are set to be proportional to the 2/3 power of the normalized dipole matrix element [29]. As for the relaxation between magnetic sublevels of the ground states, we consider two different processes, depending on the experimental conditions. One is a uniform relaxation process [24] (illustrated in Fig. 2 (a)) that is mainly caused by diffusion of Cs atoms, namely collisional relaxation between Cs and wall surfaces, and the replacement of atoms in the optical path with those in thermal equilibrium outside the optical path. The source matrix for the uniform relaxation process $\Lambda_l^{(uni)}$ is written as

$$\Lambda_l^{(uni)} = \sum_{m=1, m \neq l}^{16} \frac{1}{15} \gamma_p^{(uni)} \rho_{mm} + \sum_{n=17}^{32} \frac{T_{nl}^{2/3}}{\Sigma_{k=1}^{16} T_{nk}^{2/3}} \Gamma \rho_{nn} \quad \text{for } l = 1, \ldots, 16, \tag{12}$$

where $\gamma_p^{(uni)}$ is the decay rate of a ground state $|m\rangle$ for $m = 1, \cdots, 16$. The first and second terms represent the influxes from ground states other than itself and from the excited states, respectively. $T_{nl}$ is the normalized dipole matrix element, which satisfies $\Sigma_{l=1}^{16} T_{nl}^2 = 1$.

The other relaxation process is caused by the spin exchange collision between Cs atoms and the collision of Cs atoms with the buffer gas, where the angular momentum of the buffer gas causes a similar action to that of a random magnetic field on the Cs atoms [30, 31]. Here we represent this repopulation process as the magnetic dipole relaxation. The transition distribution in its source matrix $\Lambda_l^{(M1)}$ is proportional to the square of the Clebsch–Gordan coefficient [25].

$$\Lambda_l^{(M1)} = \sum_{m=1, m \neq l}^{16} \tilde{T}_{nl}^2 \gamma_p^{(M1)} \rho_{mm} + \sum_{n=17}^{32} \frac{T_{nl}^{2/3}}{\Sigma_{k=1}^{16} T_{nk}^{2/3}} \Gamma \rho_{nn} \quad \text{for } l = 1, \ldots, 16, \tag{13}$$

Here, $\tilde{T}_{ml}^2 = T_{ml}^2 / (1 - T_{mm}^2)$, which satisfies $\Sigma_{l=1, l \neq m}^{16} \tilde{T}_{ml}^2 = 1$. It will be necessary to estimate how these processes contribute to the relaxation in the experimental system.

The Liouville equation (9) for each matrix element is written as

$$\frac{\partial}{\partial t} \rho_{lm} = \langle l | \frac{\partial}{\partial t} \hat{\rho} | m \rangle = -\left[ \frac{\Gamma_l + \Gamma_m}{2} + i(\delta_l - \delta_m) \right] \rho_{lm} + \frac{i}{2} \sum_{u=1}^{32} (\Omega_{lu} \rho_{um} - \rho_{lu} \Omega_{um}) + \Lambda_l \delta_{lm}, \tag{12}$$

where $\delta_{lm}$ is Kronecker's symbol. These equations are rearranged as a vector matrix equation in the form $\frac{\partial}{\partial t} \boldsymbol{\rho} = M \boldsymbol{\rho}$, where $\boldsymbol{\rho}$ is a vector consisting of 1024 elements $\rho_{lm}$ and $M$ is a (1024×1024) matrix consisting of the coefficients of $\rho_{lm}$ generated from the right side of Eq. (14). We developed a computational program to calculate steady-state solutions for $\rho_{lm}$ by equating $\frac{\partial}{\partial t} \boldsymbol{\rho}$ to zero using the condition $\Sigma_{m=1}^{32} \rho_{mm} = 1$ for a closed atomic system. From Eqs. (11–14), $\Lambda_l$ should be written to satisfy the population conservation, $\sum_{m=1}^{32} \frac{\partial}{\partial t} \rho_{mm} = 0$. In this calculation of the Liouville equation with a multi-level model, we ignore other sideband lights except for the $\pm 1^{st}$ order sidebands, and we set $E_1(t) = E_2(t) = E(t)/\sqrt{2}$ for simplicity. At the steady-state condition of Eq. (14), the population of an excited state $|n\rangle$ ($n = 17, \cdots, 32$), $\rho_{nn}$, is written as

$$\rho_{nn} = -\sum_{l=1}^{16} \frac{\text{Im}(\Omega_{nl}\rho_{ln})}{\Gamma_n}. \tag{13}$$

Then, the difference between the sums of the populations of the excited states $\sum_{n=17}^{32} \rho_{nn}$ on resonance and off resonance is proportional to the experimentally observed amplitude of the CPT resonance.

.

### B. Line shape, width, and light shift

We reveal how the formulations for the line shape, width, and light shift of the CPT resonance spectrum are guided from the present multi-level atomic model. Here, we assume that the lower ground level $|g\rangle$ ($g = 1, \cdots, 7$) and the upper ground level $|e\rangle$ ($e = 8, \cdots, 16$) constitute the CPT resonance. As a necessary condition for CPT resonance to be observed, there exists at least one excited level $|i\rangle$ ($i = 17, \cdots, 32$) such that the Rabi frequency determined by the given excitation light is $\Omega_{gi} \neq 0$ and $\Omega_{ei} \neq 0$. Substituting $l = g$, $m = e$ into Eq. (13), the coherence $\rho_{ge}$ excited between the ground-state sublevels, $|g\rangle$ and $|e\rangle$, can be written as

$$\left[\frac{\Gamma_g + \Gamma_e}{2} + i(\delta_g - \delta_e)\right]\rho_{ge} = \frac{i}{2}\sum_{u=17}^{32}\left(\Omega_{gu}\rho_{ue} - \Omega_{eu}^*\rho_{gu}\right), \tag{14}$$

where $u$ refers to one of the excited levels.

Since $\Gamma_u = \Gamma \gg \Gamma_g$ or $\Gamma_e$ for any of the excited levels $|u\rangle$ ($u = 17, \cdots, 32$), we define optical decoherence as $\gamma_f = \frac{\Gamma}{2} \cong \frac{\Gamma_u + \Gamma_g}{2} \cong \frac{\Gamma_u + \Gamma_e}{2}$. $\rho_{ue}$ and $\rho_{gu}$ are then written as

$$\rho_{ue} = \frac{i}{2}\frac{\Omega_{gu}^*\rho_{ge} + \sum_{s=1,s\neq g}^{16}\Omega_{su}^*\rho_{se} - \sum_{t=17}^{32}\Omega_{et}^*\rho_{ut}}{\gamma_f + i(\delta_u - \delta_e)},$$

$$\rho_{gu} = \frac{i}{2}\frac{-\Omega_{eu}\rho_{ge} + \sum_{t=17}^{32}\Omega_{gt}\rho_{tu} - \sum_{s=1,s\neq g}^{16}\Omega_{su}\rho_{gs}}{\gamma_f - i(\delta_u - \delta_g)}. \tag{15}$$

We substitute these into the right-hand side of Eq. (16) and transfer the term containing $\rho_{ge}$ to the left-hand side. Then, $\rho_{ge}$ is solved to be

$$\rho_{ge} = \frac{-\frac{1}{4}\sum_{u=17}^{32}\left[\frac{\Omega_{gu}\left(\sum_{s=1,s\neq g}^{16}\Omega_{su}^{*}\rho_{se} - \sum_{t=17}^{32}\Omega_{et}^{*}\rho_{ut}\right)}{\gamma_f + i(\delta_u - \delta_e)} + \frac{\Omega_{eu}^{*}\left(\sum_{s=1,s\neq g}^{16}\Omega_{su}\rho_{gs} - \sum_{t=17}^{32}\Omega_{gt}\rho_{tu}\right)}{\gamma_f - i(\delta_u - \delta_g)}\right]}{\Delta_{width} + i(\delta_g - \delta_e - \Delta_{LS})} \quad (16)$$

where

$$\Delta_{width} = \frac{\Gamma_g + \Gamma_e}{2} + \frac{1}{4}\sum_{u=17}^{32}\left[\frac{|\Omega_{gu}|^2 \gamma_f}{\gamma_f^2 + (\delta_u - \delta_e)^2} + \frac{|\Omega_{ue}|^2 \gamma_f}{\gamma_f^2 + (\delta_u - \delta_g)^2}\right] \quad (17)$$

and

$$\Delta_{LS} = -\frac{1}{4}\sum_{u=17}^{32}\left[-\frac{|\Omega_{gu}|^2 (\delta_e - \delta_u)}{\gamma_f^2 + (\delta_u - \delta_e)^2} + \frac{|\Omega_{ue}|^2 (\delta_g - \delta_u)}{\gamma_f^2 + (\delta_u - \delta_g)^2}\right]. \quad (18)$$

By replacing the numerator on the right side of Eq. (18) with C, namely

$$C = -\frac{\gamma_f}{4\left[\gamma_f^2 + (\Delta_{opt} + \Delta'_{hfs}/2)^2\right]}\left[(\rho_{gg} + \rho_{ee}) + i(\rho_{gg} - \rho_{ee})\frac{\Delta_{opt} + \Delta'_{hfs}/2}{\gamma_f}\right]\sum_{u=17}^{23}\Omega_{gu}\Omega_{eu}^{*}$$
$$-\frac{\gamma_f}{4\left[\gamma_f^2 + (\Delta_{opt} - \Delta'_{hfs}/2)^2\right]}\left[(\rho_{gg} + \rho_{ee}) + i(\rho_{gg} - \rho_{ee})\frac{\Delta_{opt} - \Delta'_{hfs}/2}{\gamma_f}\right]\sum_{u=24}^{32}\Omega_{gu}\Omega_{eu}^{*}, \quad (21)$$

the real part of $\rho_{ge}$, $\text{Re}(\rho_{ge})$, which determines the shape of the CPT resonance curve in the case where $\Omega_1 \Omega_2$ is a real number, becomes

$$\text{Re}(\rho_{ge}) = -\frac{\text{Re}(C)\Delta_{width}}{\Delta_{width}^2 + (\delta_g - \delta_e - \Delta_{LS})^2} + \frac{\text{Im}(C)(\delta_g - \delta_e - \Delta_{LS})}{\Delta_{width}^2 + (\delta_g - \delta_e - \Delta_{LS})^2}. \quad (22)$$

Here, $\text{Re}(C)$ and $\text{Im}(C)$ are the real part and the imaginary part of C. The line shape of the CPT resonance as a function of $\delta_g - \delta_e$ is composed of the sum of the symmetric Lorentzian function (first term) and the antisymmetric Lorentzian function (second term), with the width $\Delta_{width}$ and the light shift $\Delta_{LS}$. This line shape matches that obtained for the three-level model, except for their amplitudes ($\text{Re}(C)$ and $\text{Im}(C)$). Thus, we can derive the line shape, width, and light shift of the CPT resonance for the multi-level atomic model. Eq. (21) is the simplified form of C in a condition where CPT resonance frequencies are resolved clearly by the magnetic field and the frequencies of the excitation lights are tuned to the excited state $6P_{1/2}$ $F' = 3$ or 4. Note that the term $\Delta_{opt} + \Delta'_{hfs}/2$ (or $\Delta_{opt} - \Delta'_{hfs}/2$) is nearly zero when the excitation lights are tuned to $F' = 3$ (or 4) levels. When the excitation lights are detuned, asymmetry in the CPT spectrum appears, as reported in [21,22,32,33]. A rigorous derivation of the CPT spectrum in the 32-level model is presented in the Appendix.

III.       EXPERIMENTAL SETUP

The experimental setup is shown in Fig. 3, which is nearly identical to the one described in our previous paper [20]. Magnetic materials were carefully removed from the vicinity of the cell to reduce the inhomogeneous magnetic field and the width of the magnetic-field-sensitive CPT resonance. The residual inhomogeneity of the magnetic field is estimated to be less than 0.3 µT. We utilized three Cs-vapor cells filled with different pressure of nitrogen buffer gas. The Cs-vapor cells were used at room temperature and not actively controlled by heating devices. The buffer gas pressure for each cell was determined from the measured value of the CPT resonance frequency shift based on the reported value of the buffer gas shift [34]. The length of the cells was 25 mm. The pressure and shape of used Cs-vapor cells are listed in Table I, together with the decay rate of the ground states $\gamma_p$ and the excited-state relaxation rate $\Gamma$, whose values are necessary to calculate the CPT resonance. $\gamma_p$ was determined experimentally from the CPT resonance width when the excitation light intensity is close to zero (See Section IV.D). We also estimated $\gamma_p$ from the Eq. (3.5.6) and Eq. (3.6.98) in ref. [31] using the pressure of the buffer gas, the configuration of the cell, the diffusion coefficient, and temperature. We confirmed that the estimated values of $\gamma_p$ were consistent with experimental values in the order of magnitudes. The excited-state relaxation rate $\Gamma$ was determined from the width (FWHM) of absorption spectrum, which results from the convolution of collision broadening and Doppler broadening [28]. The width of the Doppler broadening at room temperature is about 360 MHz. In the present paper, uncertainties are given by a standard deviation.

The Cs Cell was irradiated by a distributed Bragg reflector (DBR) laser beam, whose wavelength was tuned to the vicinity of transition from $6S_{1/2}$ to $6P_{1/2}$ $F'=3$ or 4 in the $D_1$ line of $^{133}Cs$. The laser light was modulated by an electro-optic modulator (EOM) driven with a modulation frequency $f_m$ set to around a half of hyperfine frequency splitting. The +first order and –first order sideband frequencies were used as bichromatic excitation lights $\omega_1$ and $\omega_2$, whose intensities are identical. The polarization of excitation light was changed to circular polarization or linear polarization using a quarter-wave plate. The frequency-modulated light was expanded to be a 7.2-mm diameter beam at the center of a Cs-vapor cell and passed through the cell. The intensity of laser beam used in this experiment was less than 15 µW /mm², which is at most 20% of the saturation intensity of the $D_1$ line [27].

The spectrum, amplitude and width of the CPT resonance were measured as a function of detuning frequency which is difference from the center modulation frequency of the (0,0) resonance. The measured CPT amplitudes were defined by the difference of the transmitted beam intensity between on-resonance (peak) and far-off resonance in µW/mm², which are compared to the CPT amplitude calculated using Eq. (15) with the experimental values of $\gamma_p$ and $\Gamma$.

IV. RESULTS AND DISCUSSION

A. Dependence of CPT spectrum on buffer-gas pressures

Figures 4 (a)-(c) show the experimental CPT spectra observed in three gas cells with different buffer gas pressures of 0.09, 1.35, and 11.5 kPa excited by circular polarization tuned to $F'=4$ levels. These are spectra with seven Zeeman peaks, which correspond to the $(m,m)$ resonance between two ground levels of $|F=3, m_F=m\rangle$ and $|4, m\rangle$ from $m = -3$ to $+3$, respectively. The experimental results show how large the width and the peak amplitude pattern of the CPT resonance vary with the gas pressure of the cell. Some experimental spectra show asymmetric feature, because the scan speed of the frequency is too fast. In cells with higher buffer gas pressure, the width of the CPT resonance becomes narrower. As for the amplitude distribution of the CPT spectra, the CPT resonance of (–3, –3) in the high-pressure cell becomes more significant, which may suggest that the population distribution among the ground states is different depending on the buffer gas pressure.

Warren et al. calculated the CPT spectrum of $^{87}$Rb vapor with 10 Torr (1.3 kPa) Ne buffer gas at 53°C using "the uniform relaxation" represented in Eq. (12) as a source matrix [23]. In a different approach, Matsuda et al. calculated the CPT spectrum of $^{133}$Cs vapor with 10 kPa Ne - Ar buffer gas at 80°C using "the magnetic dipole relaxation" represented in Eq. (13) [25]. The two relaxations are illustrated in Fig. 2. Here we show that such a difference in the intensity distribution will be explained by a ratio of the combination of the two relaxation processes. We assume that the relaxation rate of the ground states $\gamma_p$ is the sum of the relaxation rates of the two processes $\gamma_p^{(uni)}$ and $\gamma_p^{(M1)}$, namely, $\gamma_p = \gamma_p^{(uni)} + \gamma_p^{(M1)}$. If we define the ratio of the uniform relaxation to the total relaxation as $r$, the ratio of the magnetic dipole relaxation is $(1-r)\gamma_p$. Then, the source matrix for the total relaxation process is written as follows.

$$\Lambda_l^{(uni+M1)} = \sum_{m=1, m \neq l}^{16} \left[ \frac{1}{15} r + \tilde{T}_{nl}^2 (1-r) \right] \gamma_p \rho_{mm} + \sum_{n=17}^{32} \frac{T_{nl}^{2/3}}{\sum_{k=1}^{16} T_{nk}^{2/3}} \Gamma \rho_{nn} \quad \text{for } l = 1,\ldots,16 \qquad (23)$$

As stated in Sec. III, the values of $\gamma_p$ and $\Gamma$ in Table I were determined experimentally from the CPT spectra and the absorption spectra measured for each cell, respectively. Substituting Eq. (23) into Eq. (14), we calculated the Zeeman CPT spectrum for each cell so as to fit to the experimental results by changing $r$ as a fitting parameter. Figures 4 (d)-(f) show the CPT spectra calculated using $r$ = 1.00, 0.60, and 0.30 for the buffer gas pressures of 0.09, 1.35, and 11.5 kPa, respectively. The magnitude of the CPT amplitude for calculation is defined so that the CPT amplitude of the (0,0) resonance for 1.35 kPa is equal to the experimental one. Note that the buffer gas shift, which is clearly seen in Cell3, is not included in the present calculation. The three calculated amplitude patterns of the CPT spectrum nearly reproduce those for the corresponding experimental ones, although the calculated spectrum width for Cell1 is somewhat narrower than the experimental width. Thus, the linewidth of CPT resonance irradiated at a few μW/mm² becomes wider as $\Gamma$ is smaller. In the present calculation using multi-level model, the calculated amplitude pattern was able to reproduce the experimental pattern in each cell, but relative amplitudes between different cells are somewhat differ between them. In order to discuss more precisely, it will be necessary to consider the shape of the cell and the variation along the length of the cell.

### B. First-order Zeeman CPT spectrum

In the rest of this paper, we will discuss CPT spectrum for Cell2 (1.35 kPa). First, we study how the patterns of the Zeeman CPT spectra are related to the excitation level of $F' = 3$ or 4 and the polarization scheme of the excitation lights. In Fig. 5, blue dots indicate the observed CPT spectra and red solid lines indicate the calculated correspondent spectra for $r$ = 0.6. Then, the calculated peak amplitude of the (0, 0) resonance excited with circularly polarized lights $\sigma^- - \sigma^-$ whose frequencies are tuned to $F'=4$ is defined as being equal to the experimental one, as shown in Fig. 5(a). Figures 5(a) and 5(b) show the CPT spectra excited by $\sigma^- - \sigma^-$ tuned to $F'=4$ and $F'=3$, respectively. The (–3, –3) resonance in Fig. 5(a) produces the largest signal. In contrast, a roughly antisymmetric pattern is observed in Fig. 5(b), where the (0, 0) resonance is the largest, and the $(m,m)$ resonance with a positive value of $m$ are larger than that with a negative value. We find the patterns obtained by calculation fairly match the experimental patterns, except that the observed $(-m,-m)$ resonances become smaller than the calculated results as the value of $m$ increases. This is presumably due to the spatial inhomogeneity of the magnetic field in the experimental system.

The amplitude of the Zeeman CPT spectra is determined by the sum of populations of the ground states related to the CPT resonance, and the product of two Rabi frequencies responsible to the $\Lambda$ scheme, as shown in Eq. (21). For each resonance signal excited to $F'=4$ and 3 levels, the former and latter are respectively shown by blue triangles and red stars in Fig. 6(a) and (b). Though the distribution of the sum of the populations is similar regardless of $F'=4$ and 3 levels, the pattern of the product of Rabi frequencies is different for the two excitations depending on the values of their Clebsch-Gordan coefficients. Here, we note that the (–3, –3) resonance excited to $F'=3$ is observed in both the numerical and experimental results even though the product of the Rabi frequencies is zero, which is due to the contribution from the detuned excitation to $F'=4$ levels.

We also calculated the Zeeman CPT spectra excited with the Lin ∥ Lin polarization scheme to $F'=3$ levels, and with the Lin ⊥ Lin polarization scheme to $F'=4$ levels. They are shown by red solid lines in Fig. 5(c) and 5(d), respectively, with relative values to the calculated peak amplitude of the (0, 0) resonance excited by $\sigma^- - \sigma^-$ tuned to $F'=4$. In Fig. 5(c), the present experimental result is also overlaid with blue dots. The calculated pattern is almost identical to that observed experimentally. The pattern of Fig. 5(d) was already shown by the experiment with the push-pull optical pumping [17]. We summarized the measured and calculated peak amplitudes of the CPT resonance excited with different schemes in Table II. The calculated ratios are in fairly good agreement with the experimental ratios. We confirm using the present multi-level model that the Lin ⊥ Lin polarization scheme produces the largest amplitude of the (0, 0) CPT resonance in the $D_1$ transition of $^{133}$Cs atoms.

### C. Second-order Zeeman CPT spectrum

Depending on the strength of the quantized magnetic field, each Zeeman CPT spectrum is split into three $\Lambda$-scheme resonances of $(m,m)$, $(m-1, m+1)$, and $(m+1, m-1)$ due to the second order Zeeman effect. The resonances of $(m-1, m+1)$ and $(m+1, m-1)$ via $m'_F = m$ are created by the excitation scheme of the linear-linear polarization, whereas the $(m,m)$ resonance is created by two $\sigma^+$ polarizations via $m'_F = m+1$, or two $\sigma^-$ polarizations via $m'_F = m-1$. For such a double-$\Lambda$ scheme of the $(m,m)$ resonance generated by the linear-linear polarization, Liu et al., derived the condition where a dark state common to the two $\Lambda$ schemes exists [17], as follows.

$$e^{2i\theta} = \frac{\Omega_{2s't}^{\sigma+} \Omega_{1st'}^{\sigma-}}{\Omega_{1st}^{\sigma+} \Omega_{2s't'}^{\sigma-}} \tag{19}$$

Here, we define level $|s\rangle$ as $F=3$ and $m_F = m$, $|s'\rangle$ as $F=4$ and $m_F = m$, $|t\rangle$ as $F'=3$ and $m_F = m+1$, $|t'\rangle$ as $F' = 3$, and $m_F = m-1$. $\theta$ is the angle between two electric fields. Conversely, when the dark state for a transition is the bright state for the other transition, the two $\Lambda$ schemes act to weaken each other. The condition is as follows.

$$e^{2i\theta} = -\frac{\Omega_{1st}^{\sigma+}\Omega_{1st'}^{\sigma-}}{\Omega_{2s't}^{\sigma+}\Omega_{2s't'}^{\sigma-}} \quad (20)$$

In the case of $m=0$ in the $D_1$ transition of $^{133}$Cs, Eq. (24) becomes $e^{2i\theta} = -1$. Therefore, in the Lin $\perp$ Lin (or push-pull) scheme with $\theta = \pi/2$, a common dark state exists and the (0, 0) CPT resonance occurs, along with (–1, 1) and (1, –1) [17]. In contrast, in the Lin || Lin scheme with $\theta = 0$, no (0, 0) CPT resonance occurs, since the dark state for one $\Lambda$ scheme becomes the bright state for the other. Note that in the case of $m \neq 0$, Eq. (25) is still satisfied in the Lin || Lin scheme with $\theta = 0$ and $E_1 = E_2$.

We measured in precisely the second order Zeeman splitting of 7 CPT resonances (Fig. 5(c)) observed in the Lin || Lin scheme. The blue dots in Fig. 7 indicate experimental spectra in the vicinity of the $(m,m)$ resonances for $m = -3, \cdots, +3$ excited to $F' = 3$ levels with the Lin || Lin polarization. Except for $m=0$, we can clearly observe the $(m,m)$ resonances. The amplitude of the $(m,m)$ resonance increases as the absolute value of $m$ increases. We also calculated the CPT spectrum using the multi-level model with the normalization so that the calculated peak amplitude of the (–1, 1) resonance is identical to the experimental one. The calculated signal is overlaid by red solid lines in Fig. 7, and we find that spectra of the (–1, 1) and (1, –1) resonances agree fairly well the experimental ones. As the value of $m$ increases, the peak amplitude in the experimental spectra becomes smaller than the calculated ones. As we already stated, it occurs due to the inhomogeneity of the magnetic field, which is estimated to be less than 0.3 μT. Furthermore, we confirm that the $(m,m)$ resonances, except for $m=0$, occur in the calculated spectrum. The fact that the $(m,m)$ resonance except for $m=0$ is not forbidden in experiment and calculation suggests that an additional condition is required in the double-$\Lambda$ scheme to suppress the $(m,m)$ resonance. We explain this as follows.

As written in Eq. (15), a measure of transparency is $\sum_{n=17}^{32}\sum_{l=1}^{16} \text{Im}(\Omega_{nl}\rho_{ln})$. Here, we consider a double-$\Lambda$ scheme composed of the ground states pair, $|g\rangle$ and $|e\rangle$, and the excited states associated with $\sigma^+$ and $\sigma^-$ polarizations, $|n^+\rangle$ and $|n^-\rangle$. The contribution of this double-$\Lambda$ scheme to transmittance is given by the following.

$$\text{Im}\left(\Omega_{n^+g}\rho_{gn^+} + \Omega_{n^-g}\rho_{gn^-}\right) = \text{Im}\left[i\Omega_{n^+g}\frac{\sum_{v=17}^{32}\Omega_{gv}\rho_{vn^+} - \sum_{u=1}^{16}\Omega_{un^+}\rho_{gu}}{\Gamma_g + \Gamma_{n^+} + 2i(\delta_g - \delta_{n^+})} + i\Omega_{n^-g}\frac{\sum_{v=17}^{32}\Omega_{gv}\rho_{vn^-} - \sum_{u=1}^{16}\Omega_{un^-}\rho_{gu}}{\Gamma_g + \Gamma_{n^-} + 2i(\delta_g - \delta_{n^-})}\right]$$
$$\cong -\frac{1}{4\gamma_f}\left\{\left(|\Omega_{gn^+}|^2 + |\Omega_{gn^-}|^2\right)\rho_{gg} + \text{Re}\left[\left(\Omega_{gn^+}^*\Omega_{en^+} + \Omega_{gn^-}^*\Omega_{en^-}\right)\rho_{ge}\right]\right\} \quad (21)$$

Here, we assume the frequencies of the excitation lights are tuned to both the $n^+$ and $n^-$ levels, such that $|\delta_g - \delta_{n^\pm}| \ll \gamma_f$. Similarly,

$$\text{Im}\left(\Omega_{n^+e}\rho_{en^+} + \Omega_{n^-e}\rho_{en^-}\right) \cong -\frac{1}{4\gamma_f}\left\{\left(|\Omega_{en^+}|^2 + |\Omega_{en^-}|^2\right)\rho_{ee} + \text{Re}\left[\left(\Omega_{gn^+}^*\Omega_{en^+} + \Omega_{gn^-}^*\Omega_{en^-}\right)\rho_{ge}\right]\right\}. \quad (22)$$

The term associated with $\rho_{gg}$ or $\rho_{ee}$ is the contribution of one-photon absorption. From Eqs. (21–22), the term associated with $\rho_{ge}$ is written as follows.

$$\text{Re}\left[\left(\Omega^*_{gn^+}\Omega_{en^+}+\Omega^*_{gn^-}\Omega_{en^-}\right)\rho_{ge}\right] = \frac{\text{Re}\left[\left(\Omega^*_{gn^+}\Omega_{en^+}+\Omega^*_{gn^-}\Omega_{en^-}\right)C\right]\Delta_{width}}{\Delta_{width}^2+\left(\delta_g-\delta_e-\Delta_{LS}\right)^2} + \frac{\text{Im}\left[\left(\Omega^*_{gn^+}\Omega_{en^+}+\Omega^*_{gn^-}\Omega_{en^-}\right)C\right]\left(\delta_g-\delta_e-\Delta_{LS}\right)}{\Delta_{width}^2+\left(\delta_g-\delta_e-\Delta_{LS}\right)^2}$$

$$\cong -\frac{\rho_{gg}+\rho_{ee}}{4\gamma_f}\frac{\Delta_{width}}{\Delta_{width}^2+\left(\delta_g-\delta_e-\Delta_{LS}\right)^2}\left|\Omega^*_{gn^+}\Omega_{en^+}+\Omega^*_{gn^-}\Omega_{en^-}\right|^2$$

(23)

Therefore, the amplitude of the CPT resonance depends on the square of $\Omega^*_{gn^+}\Omega_{en^+}+\Omega^*_{gn^-}\Omega_{en^-} \equiv \Omega^{\sigma+}_{1st}\Omega^{*\sigma+}_{2s't}+\Omega^{\sigma-}_{1st'}\Omega^{*\sigma-}_{2s't'}$.

This value is 0 for $m=0$ but is proportional to 0.0065, 0.021, and 0.027 for $m=\pm1, \pm2, \pm3$, respectively. Thus, in order to prohibit the $(m,m)$ CPT resonance excited to $F'=3$ levels with the linear-linear polarization in case of the double-$\Lambda$ scheme, $\sum_{u=17}^{23}\Omega_{gu}\Omega^*_{eu}=0$ is required in addition to the conditions of $\theta=0$ and $E_1=E_2$. This condition is equivalent to

$$\Omega^{\sigma+}_{1st}\Omega^{*\sigma+}_{2s't} = -\Omega^{\sigma-}_{1st'}\Omega^{*\sigma-}_{2s't'}.$$

(24).

### D. Dependence of CPT resonance on excitation intensity

In our previous paper [20], we showed that the amplitude of the (–1, 1) and (1, –1) resonances excited with the Lin || Lin polarization increases approximately in proportion to the excitation intensity, while the amplitude of the (0, 0) resonance excited with $\sigma^--\sigma^-$ polarization moderately saturates. We concluded that the former can be described by the simple three-level model and the latter by the four-level model with a trap state. However, such behaviors should be explained by one complete equation that considers all sublevels related to the $D_1$ line of $^{133}\text{Cs}$. In this section, we compare the experimental results of the dependency of the (–1, 1) and the (0, 0) resonances on excitation intensity with the calculated ones using the present multi-level model.

The width and the amplitude of the (0, 0) resonance excited with $\sigma^--\sigma^-$ polarization tuned to $F'=3$ and $F'=4$ levels and those of the (–1, 1) resonance excited with the Lin || Lin polarization tuned to $F'=3$ and $F'=4$ levels were measured for Cell2 ($N_2$: 1.35 kPa) and plotted as a function of the excitation intensity in Fig. 8, by green (light gray) circles, blue (gray) squares, red (light gray) triangles and magenta (gray) stars, respectively. Figure 8(a) shows the measured widths of those resonances as a function of the excitation intensity, together with the calculated widths using the multi-level model with the values of $\Gamma/2\pi=0.51$ GHz and $\gamma_p/2\pi=0.107$ kHz. The pink (light gray)-shaded area of the calculated width for excitation with the Lin || Lin polarization tuned to $F'=3$ shows the uncertainty of ±4 %. The calculated widths for excitation with the Lin || Lin polarization are in good agreement with the experimental widths. This fact was already ascertained in the spectra of the (–1, 1) and (1, –1) resonance of Fig. 7(d). On the other hand, the calculated widths for excitation with the $\sigma^--\sigma^-$ polarization are wider than the experimental width by three times the uncertainty, however, the tendency of the width against the excitation intensity is similar. It is found that the width excited by the Lin || Lin polarization is wider than that excited by the $\sigma^--\sigma^-$ polarization and the width excited to $F'=4$ is wider than that excited to $F'=3$.

According to Eq. (19), the width of CPT resonance becomes $\gamma_p$ at the excitation intensity of almost zero, but it increases approximately with the slope due to the reciprocal of $\Gamma$ as intensity increases, as stated in Section IV.A.

Thus, the experimental decay rate of the ground states $\gamma_p/2\pi$ = 0.107±0.006 kHz was obtained from the CPT resonance width when the excitation light intensity is close to zero.

Figure 8(b) shows the comparison of the experimental and calculated amplitudes of CPT resonance. The amplitudes excited by the Lin||Lin polarization tuned to F'=3 and F'=4 increase proportionally with intensity, as shown by red (light gray) triangles and magenta (gray) stars, respectively. The magnitude of the experimental CPT amplitude is order of nW/mm$^2$ for incident light of a few µW/mm$^2$. The calculated amplitude excited with the Lin || Lin scheme to $F'=3$ is fitted to the experimental one, as shown in red line agrees quite well with the experimental values within the uncertainty (±20%) of the calculated amplitude, which are shown by the pink (light gray)-shaded area. Then the calculated amplitude excited by the Lin||Lin to F'=4 in magenta line agrees with the experimental values similarly. Contrary to this, the calculated amplitude excited with $\sigma^- - \sigma^-$ scheme was about twice as the calculated amplitudes excited with the Lin || Lin scheme. Therefore, the calculated amplitude excited with $\sigma^- - \sigma^-$ scheme to $F'=4$ is also fitted to the measurement data, as shown in blue line. Then, we confirm that the relative relationships between $F'=4$ and $F'=3$ are in good agreement with the experimental values. Similar to our previous results [20], the amplitude of the CPT resonance excited with $\sigma^- - \sigma^-$ polarization saturates as excitation intensity increases. On the other hand, that excited with Lin || Lin polarization is increasing in proportion to the excitation intensity because the excitation intensity is less than the saturation intensity of $D_1$ line. For reference, we note that the absorbed intensity of cell irradiated at $I$ = 2.5 µW/mm$^2$ was 0.24 µW/mm$^2$ and 0.36 µW/mm$^2$ for excitation by the $\sigma^- - \sigma^-$ polarization tuned to $F'=4$ and by the Lin || Lin polarization tuned to $F'=3$, respectively.

To investigate the difference between dependencies on the excitation intensity for $\sigma^- - \sigma^-$ excitation and Lin || Lin excitation, we calculated the population of trap states using the 32-level model. The state $|8\rangle$ in Fig. 1 is a trap state for the excitation to $F'=4$ levels, and states $|1\rangle$, $|8\rangle$, and $|9\rangle$ are trap states for the excitation to $F'=3$ levels. The populations of the trap states for several excitation schemes are shown in Fig. 9 as a function of the excitation intensity. In the case of no excitation light, the populations of the ground states are thermally equivalent, namely the population of trap state is 19% for $F'=3$ and 6.3% for $F'=4$. For $\sigma^- - \sigma^-$ excitation, the population of the trap state increases as the intensity increases and reaches to 60% at more than 3 µW/mm$^2$. In contrast, that for Lin || Lin excitation does not depend on the excitation intensity and is almost constant. These findings clarify that the CPT amplitude for $\sigma^- - \sigma^-$ excitation saturates while the CPT amplitude for the Lin || Lin excitation increases in proportion to the excitation intensity. The population of the trap state for the Lin ⊥ Lin excitation to $F'=4$ levels versus excitation intensity is almost same as that for Lin || Lin excitation to the $F'=4$ levels. The reason why the calculated CPT amplitude for $\sigma^- - \sigma^-$ excitation deviate from the experimental one to some extent may be that it reflects real population of the trap state.

Thus, the phenomena explained separately using the separate equations of three level system and four level system in the previous paper can be explained uniformly by calculation using the equations of the muti-level model. As stated above, many phenomena can be calculated using the present multi-level model. However, there is not a complete agreement between the experimental data and the calculated results using the present model. To There are several things that need to consider for improvement. It would be necessary to add spatial characteristics that reflect the Gaussian intensity distribution of the excitation light, the attenuation of light in the thickness direction of the gas cell, saturation effect of the light, the spatial non-uniformity of the magnetic field, and so on. Although we used a frequency-modulated light for the excitation of Cs atoms in the present experiment, we took into consideration +first sideband and –first sideband frequencies as bichromatic light in the present model. More precisely, the interactions with a carrier frequency and other higher-order sideband frequencies should be included in the model.

## V. CONCLUSION

We constructed a computational multi-level atomic model of the Liouville density-matrix equation to investigate the CPT resonances excited by the bichromatic lights of various excitation schemes between the ground hyperfine levels. The model contains 32 Zeeman sublevels on the $D_1$ line of $^{133}Cs$ atoms. We also derived formulations for the line shape, width, and light shift of the CPT resonance spectrum analytically from the present multi-level atomic model. By calculating the model numerically with the experimentally determined decay rates of the ground and the excited states, the amplitude and shape of the CPT resonance were obtained for different excitations by circular or linear polarization. We confirmed that the calculations accurately reproduced the experimental spectra observed in Cs vapor cells and elucidated the mechanism underlying various characteristics. Specifically, we found that the Zeeman CPT spectra with different buffer gas pressures vary due to the relaxation process. Calculation using the present model was confirmed that the observed pattern of the first order Zeeman CPT spectra varies depending on the excitation scheme of polarization and the excitation level of $F' = 3$ or 4. The $(m,m)$ CPT resonance (except for (0, 0)) appears in the second order Zeeman CPT spectrum in the Lin ∥ Lin excitation, which demonstrates the need for an additional condition to prohibit the CPT resonance. We also clarified that the amplitude for $\sigma^- - \sigma^-$ excitation saturates while that for Lin ∥ Lin excitation increases in proportion to the excitation intensity coincident with a variation of the population of the trap state. The qualitative saturation in $\sigma^- - \sigma^-$ excitation can be reproduced by the calculation with 4-level model. However, with the 32-level model, we can further reproduce the relative difference between the excited levels $F' = 3$ and $F' = 4$, and the difference between other excitation schemes, such as the Lin ⊥ Lin polarization scheme. Thus, using the present 32-level model, we can know the features such as width, amplitude, and symmetry of the CPT resonance under various specifications.

These findings indicate that our computational multi-level model can help clarify the phenomena of the CPT resonance and promote the development of miniature atomic clock devices. To further improve the multi-level atomic model, it would be necessary to add spatial characteristics of the Gaussian intensity distribution of the excitation light, the attenuation of light in the thickness direction, the saturation effect of 1-photon absorption, the spatial non-uniformity of the magnetic field, the interaction with extra sideband lights, and so on.

APPENDIX: CPT spectrum in 32-level model

Here we consider the shape of the CPT spectrum using the 32-level model. First, we introduce shorthand symbols to represent symmetric and asymmetric Lorentz functions, namely, $S(x,w) = \frac{w}{x^2+w^2}$ and $A(x,w) = \frac{x}{x^2+w^2}$ $= S(x,w)*x/w$. We can find $(w \pm ix)^{-1} = S(x,w) \mp iA(x,w)$. From the Liouville equation in Eq. (14), for indices $g = 1,\cdots,7$ and $i = 17,\cdots,32$, $\text{Im}(\Omega_{gi}\rho_{ig})$ is written as follows.

$$\text{Im}(\Omega_{gi}\rho_{ig}) = \frac{S(\delta_i - \delta_g, \gamma_f)}{2}\left[|\Omega_{gi}|^2(\rho_{gg} - \rho_{ii}) + \text{Re}\left(\sum_{v=8}^{16}\Omega_{vi}\Omega_{ig}\rho_{gv}\right) - \frac{\delta_i - \delta_g}{\gamma_f}\text{Im}\left(\sum_{v=8}^{16}\Omega_{vi}\Omega_{ig}\rho_{gv}\right)\right] \quad (A1)$$

For simplicity, we ignore the coherence between any two magnetic sublevels in the same hyperfine state. In a similar manner, for indices $e = 8,\cdots,16$ and $i = 17,\cdots,32$, $\text{Im}(\Omega_{ei}\rho_{ie})$ is written as follows.

$$\text{Im}(\Omega_{ei}\rho_{ie}) = \frac{S(\delta_i - \delta_e, \gamma_f)}{2}\left[|\Omega_{ei}|^2(\rho_{ee} - \rho_{ii}) + \text{Re}\left(\sum_{v=1}^{7}\Omega_{ei}\Omega_{iv}\rho_{ve}\right) - \frac{\delta_i - \delta_e}{\gamma_f}\text{Im}\left(\sum_{v=1}^{7}\Omega_{ei}\Omega_{iv}\rho_{ve}\right)\right] \quad (A2)$$

Here, we define transmittance $T$ as $T = 1 - \alpha\sum_{n=17}^{32}\sum_{l=1}^{16}\rho_{ii}\text{Im}(\Omega_{ln}\rho_{nl})$, $\alpha = \frac{N_{atom}\hbar\omega\Gamma}{(I_1+I_2)S_{beam}}$. $N_{atom}$, $\hbar\omega$, and $S_{beam}$ are the number of interacting atoms, the energy of a photon of the excitation light, and the cross section of the beam, respectively. Thus, $T$ can be written as

$$T = 1 - \sum_{i'=17}^{32}\sum_{l'=1}^{16}F_1(l',i') + \sum_{g'=1}^{7}\sum_{e'=8}^{16}F_2(g',e') \quad (A3)$$

$$F_1(l,i) = \frac{\alpha}{2\Gamma}S(\delta_i,\gamma_f)|\Omega_{li}|^2(\rho_{ll}-\rho_{ii}), \quad F_2(g,e) = -\frac{\alpha}{2\Gamma}\sum_{i=17}^{32}S(\delta_i,\gamma_f)\text{Re}\left[\Omega_{ei}\Omega_{ig}\rho_{ge}\right], \quad (A4)$$

where $S(\delta_i - \delta_g, \gamma_f) \cong S(\delta_i - \delta_e, \gamma_f)$ is replaced with $S(\delta_i, \gamma_f)$. We also assume $|\delta_g - \delta_e| \ll \gamma_f$, as this is satisfied in typical situations for CPT resonance measurements. $F_1(l,i)$ represents the one-photon absorption from $|l\rangle$ to $|i\rangle$. $F_2(g,e)$ represents the CPT spectrum corresponding to the resonance state of $|g\rangle$ and $|e\rangle$. We now focus on the details of the CPT spectrum, $F_2(g,e)$. From Eq. (18),

$$-\text{Re}(\Omega_{ig}\Omega_{ei}\rho_{ge}) = S(\Delta_{ge-LS},\Delta_{ge-w})\text{Re}(M) + A(\Delta_{ge-LS},\Delta_{ge-w})\text{Im}(M) \quad (A5)$$

$$M \cong \frac{1}{4}\sum_{u=17}^{32}S(\delta_u,\gamma_f)\Omega_{ig}\Omega_{ei}\Omega_{gu}\Omega_{ue}\left[\left(1-i\frac{\delta_u-\delta_e}{\gamma_f}\right)\rho_{ee} + \left(1+i\frac{\delta_u-\delta_g}{\gamma_f}\right)\rho_{gg}\right] \quad (A6)$$

Here, we denote $\Delta_{ge-LS} = \delta_g - \delta_e - \Delta_{LS}$ and $\Delta_{ge-w} = \Delta_{width}$. In cases of $\sigma - \sigma$, Lin $\|$ Lin, or Lin $\perp$ Lin excitations, one of the real or imaginary parts of $\Omega_{ig}\Omega_{ei}$ is zero for any values of $g = 1,\cdots,7$, $e = 8,\cdots,16$, and $i = 17,\cdots,32$, which can be confirmed from Eq. (8). Thus, $\Omega_{ig}\Omega_{ei}\Omega_{gu}\Omega_{ue}$ is a real value.

Since the Zeeman shift is negligibly small relative to the hyperfine splitting between $6P_{1/2}$ $F'=3$ and 4 levels ($\Delta'_{hfs}$ in Fig. 1), we can rewrite $S(\delta_u,\gamma_f) = S\left(\delta_{F'=3} = \Delta_{opt} + \dfrac{\Delta'_{hfs}}{2}, \gamma_f\right)$ for $u=17,\cdots,23$ and $S(\delta_u,\gamma_f) = S\left(\delta_{F'=4} = \Delta_{opt} - \dfrac{\Delta'_{hfs}}{2}, \gamma_f\right)$ for $u=24,\cdots,32$. When the excitation lights are tuned closely to either of the $F'=3$ or 4 levels and ${\Delta'_{hfs}}^2 \gg \gamma_f^2$ is satisfied, we can assume $S(\delta_{F'=4},\gamma_f) \ll S(\delta_{F'=3},\gamma_f)$ or $S(\delta_{F'=3},\gamma_f) \ll S(\delta_{F'=4},\gamma_f)$, respectively. Finally, we obtain the CPT spectrum in the case of excitation to $F'=3$ and 4 levels ($F_2^{F'=3}(g,e)$ and $F_2^{F'=4}(g,e)$, respectively) as follows.

$$F_2^{F'=3}(g,e) = \frac{\alpha}{8\Gamma} S(\delta_{F'=3},\gamma_f)^2 \left|\sum_{i=17}^{23} \Omega_{ig}\Omega_{ei}\right|^2 \left[ S(\Delta_{ge-LS},\Delta_{ge-w})(\rho_{gg}+\rho_{ee}) + A(\Delta_{ge-LS},\Delta_{ge-w})\frac{\delta_{F'=3}(\rho_{gg}-\rho_{ee})-\delta_g\rho_{gg}+\delta_e\rho_{ee}}{\gamma_f} \right] \quad (A7)$$

$$F_2^{F'=4}(g,e) = \frac{\alpha}{8\Gamma} S(\delta_{F'=4},\gamma_f)^2 \left|\sum_{i=24}^{32} \Omega_{ig}\Omega_{ei}\right|^2 \left[ S(\Delta_{ge-LS},\Delta_{ge-w})(\rho_{gg}+\rho_{ee}) + A(\Delta_{ge-LS},\Delta_{ge-w})\frac{\delta_{F'=4}(\rho_{gg}-\rho_{ee})-\delta_g\rho_{gg}+\delta_e\rho_{ee}}{\gamma_f} \right] \quad (A8)$$

These are written as the sum of one symmetric and one asymmetric Lorentz function, whose width is $\Delta_{ge-w}$ and whose center is $\Delta_{ge-LS}$.

## ACKNOWLEDGEMENTS

This study was supported by Innovative Science and Technology Initiative for Security Grant Number JPJ004596, ATLA, Japan. K.M. thanks Dr. Yuichiro Yano of National Institute of Information and Communications Technology, for his kind introduction of his primitive work.

TABLE I. The pressure and shape of the gas cells used in the present experiments with the decay rate of the ground states $\gamma_p$ and the decay rate of the excited states $\Gamma$, which are determined experimentally. The length of the cells are 25 mm. Uncertainties are given by a standard deviation.

| Cells | Cell1 | Cell2 | Cell3 |
|---|---|---|---|
| $N_2$ pressure / kPa | 0.09±0.01 | 1.35±0.05 | 11.5±0.4 |
| Shape | Square | Circle | Square |
| Cross section | 20×20 mm² | 10 mm radius | 20×20 mm² |
| $\gamma_p / 2\pi$ / kHz | 24.5±0.8 | 0.107±0.006 | 0.081±0.006 |
| $\Gamma / 2\pi$ / GHz | 0.38±0.03 | 0.51±0.03 | 1.69±0.18 |

TABLE II. The amplitude ratio of the CPT resonances excited with different schemes at $I_1 = I_2 = 6.6\,\mu W / mm^2$.

| Polarization | $\sigma^- - \sigma^-$ | $\sigma^- - \sigma^-$ | Lin ∥ Lin | Lin ⊥ Lin |
|---|---|---|---|---|
| Excitation levels | $F' = 4$ | $F' = 3$ | $F' = 3$ | $F' = 4$ |
| CPT resonance | (0, 0) | (0, 0) | (–1, 1) and (1, –1) | (0, 0) |
| Relative amplitude of experiment | 1.0 | 0.44 | 3.6 | - |
| Relative amplitude of calculation | 1.0 | 0.47 | 3.2 | 17 |

Figure Captions

FIG. 1 (Color online) Hyperfine Zeeman structure of the $D_1$ line of $^{133}$Cs. Values are expressed in the unit of angular frequency. $|1\rangle - |7\rangle$, $|8\rangle - |16\rangle$, $|17\rangle - |23\rangle$, and $|24\rangle - |32\rangle$ are the magnetic sublevels of $6S_{1/2}\ F = 3$, $6S_{1/2}\ F = 4$, $6P_{1/2}\ F' = 3$, and $6P_{1/2}\ F' = 4$, respectively, labeled in order of magnetic quantum number $m_F$ from $-F$ to $F$. $\omega_g^0$ and $\omega_e^0$ are unperturbed energies of $6S_{1/2}\ F = 3$ and $6S_{1/2}\ F = 4$, respectively. $\omega_{i0}$ is the medium energy between the unperturbed energies of $6P_{1/2}\ F' = 3$ and $6P_{1/2}\ F' = 4$. $\Delta_{hfs}$ and $\Delta'_{hfs}$ are the hyperfine splitting energies of $6S_{1/2}$ and $6P_{1/2}$, respectively. $\omega_1$ and $\omega_2$ are frequencies of the excitation lights in a unit of angular frequency. $\Delta_R$ and $\Delta_{opt}$ are the Raman detuning and the common detuning, respectively, which are defined in the manuscript. $\Gamma$ and $\gamma_p$ are the decay rate of $6P_{1/2}$ (excited) states and $6S_{1/2}$ (ground) states, respectively.

FIG. 2 (Color online) Schematic diagrams of the relaxation mechanism between ground levels. (a) The uniform relaxation process. (b) The magnetic dipole relaxation process.

FIG. 3 (Color online) Experimental setup. VOA: variable optical attenuator, HWP: half-wave plate, PBS: polarizing beam splitter, QWP: quarter-wave plate, PD: photodetector.

FIG. 4 (Color online) Zeeman CPT spectra excited with circularly polarized lights $\sigma^- - \sigma^-$ tuned to the $F' = 4$ levels at $B = 22.7\ \mu T$ observed in three cells at different $N_2$ pressures. The excitation intensities are $I_1 = I_2 = 6.6\ \mu W / mm^2$. The scan speed of the frequency was 600 kHz/sec. Experimental spectra in (a) Cell1; 0.09 kPa, (b) Cell2; 1.35 kPa, and (c) Cell3; 11.5 kPa. Calculated spectra with the corresponding parameters for each gas cell of (d) Cell1, (e) Cell2, and (f) Cell3. $r$ is the ratio of the uniform relaxation to the total relaxation.

FIG. 5 (Color online) Zeeman CPT spectra in Cell2 at $B = 22.7\ \mu T$. $N_2$ pressure is 1.35 kPa (Cell2), and the excitation intensities are $I_1 = I_2 = 6.6\ \mu W / mm^2$. The scan speed of the frequency was 600 kHz/sec. Experimental (blue (gray) dot) and calculated (red (light gray) line) CPT spectra excited with circularly polarized light $\sigma^- - \sigma^-$ tuned to (a) $F' = 4$ and (b) $F' = 3$ levels, (c) CPT spectra excited with Lin ∥ Lin schemes tuned to $F' = 3$ level, and (d) Lin ⊥ Lin scheme tuned to $F' = 4$ level. Calculations are executed with $r$ = 0.6, $\Gamma / 2\pi$ = 0.51 GHz, and $\gamma_p / 2\pi$ = 0.107 kHz at $B = 139\ \mu T$. The amplitude of the calculated (0,0) CPT resonance excited with circularly polarized light $\sigma^- - \sigma^-$ tuned to $F' = 4$ is normalized to the experimental one.

FIG. 6 (Color online) The population of two ground states (blue (gray) triangle) and the product of Rabi frequencies (red (light gray) star) for the CPT resonance tuned to (a) $F' = 4$ and (b) $F' = 3$ levels. The indices $g$, $e$, and $i$ in the product of Rabi frequencies indicate the two ground levels and the excited level of the corresponding Λ transition.

FIG. 7 (Color online) Calculated (red (light gray) line) and experimental (navy (gray) dot) CPT spectra in Cell2

near the $(m, m)$ resonance for $m = -3, \cdots, +3$, by Lin || Lin excitation tuned to $F' = 3$ levels at $B = 285\,\mu\text{T}$. The excitation intensities are $I_1 = I_2 = 1.2\,\mu\text{W}/\text{mm}^2$. The scan speed of the frequency was 80 kHz/sec. Calculations are executed with $r = 0.6$, $\Gamma/2\pi = 0.51$ GHz, and $\gamma_p/2\pi = 0.107$ kHz. The amplitude of the calculated CPT spectra are shown with the normalization so that the calculated peak amplitude of the (–1, 1) resonance is identical to the experimental one.

FIG. 8 (Color online) Measured and calculated widths and amplitudes of the (0, 0) or (–1, 1) resonance for different polarization and excitation levels as a function of the excitation intensity $I$ at $B = 139\,\mu\text{T}$. The scan speed of the frequency was 100 kHz/sec. $N_2$ pressures is 1.35 kPa (Cell2). Calculations are executed with $r = 0.6$, $\Gamma/2\pi = 0.51$ GHz, and $\gamma_p/2\pi = 0.107$ kHz. (a) The widths and (b) the intensities of the CPT resonances as a function of the excitation intensity. The calculated amplitudes excited with $\sigma^- - \sigma^-$ scheme is normalized to the experimentally measured amplitude tuned to $F' = 4$ levels, while the calculated amplitudes excited with the Lin || Lin scheme is normalized to the experimentally measured amplitude tuned to $F' = 3$ levels. Green (light gray) circle (measured) and dotted line (calculated): the (0, 0) resonance excited with $\sigma^- - \sigma^-$ scheme tuned to $F' = 3$ levels. Blue (gray) square (measured) and dashed line (calculated): the (0, 0) resonance excited with $\sigma^- - \sigma^-$ scheme tuned to $F' = 4$ levels. Red (light gray) triangle (measured) and solid line (calculated): the (–1, 1) resonance excited with Lin || Lin scheme tuned to $F' = 3$ levels. Magenta (gray) star (measured) and dash-dot line (calculated): the (–1, 1) resonance excited with Lin || Lin scheme tuned to $F' = 4$ levels.

FIG. 9 (Color online) The calculated population of trap states for several excitation schemes as a function of excitation intensity. Calculations are executed with $r = 0.6$, $\Gamma/2\pi = 0.51$ GHz, and $\gamma_p/2\pi = 0.107$ kHz at $B = 139\,\mu\text{T}$, which are optimal values for Cell2. Green (light gray) dotted line: the (0, 0) resonance excited with $\sigma^- - \sigma^-$ scheme tuned to $F' = 3$ levels. Blue (gray) dashed line: the (0, 0) resonance excited with $\sigma^- - \sigma^-$ scheme tuned to $F' = 4$ levels. Red (light gray) solid line: the (–1, 1) resonance excited with Lin || Lin scheme tuned to $F' = 3$ levels. Magenta (gray) dash-dot line: the (–1, 1) resonance excited with Lin || Lin scheme tuned to $F' = 4$ levels.

FIG. 1

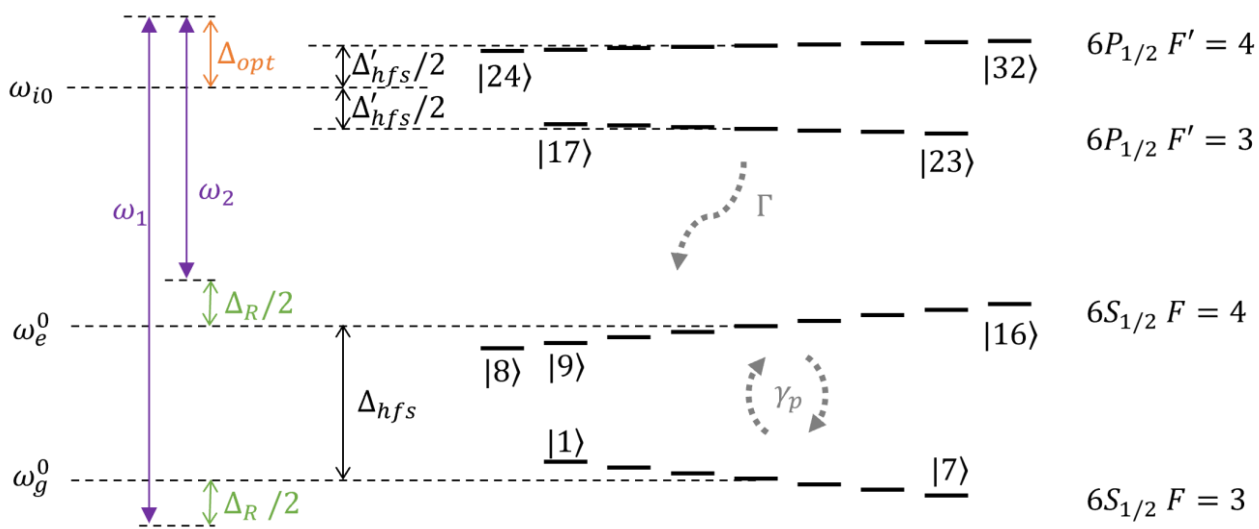

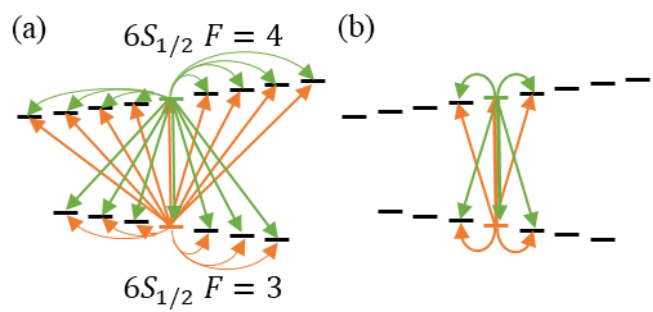

FIG. 2

FIG3

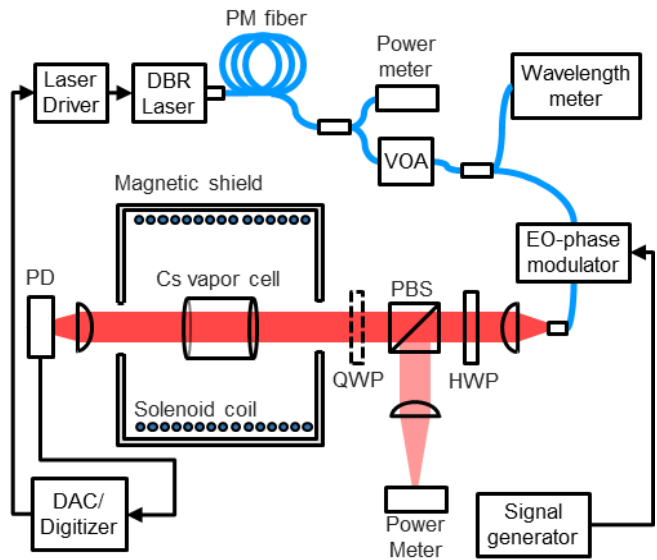

FIG. 4

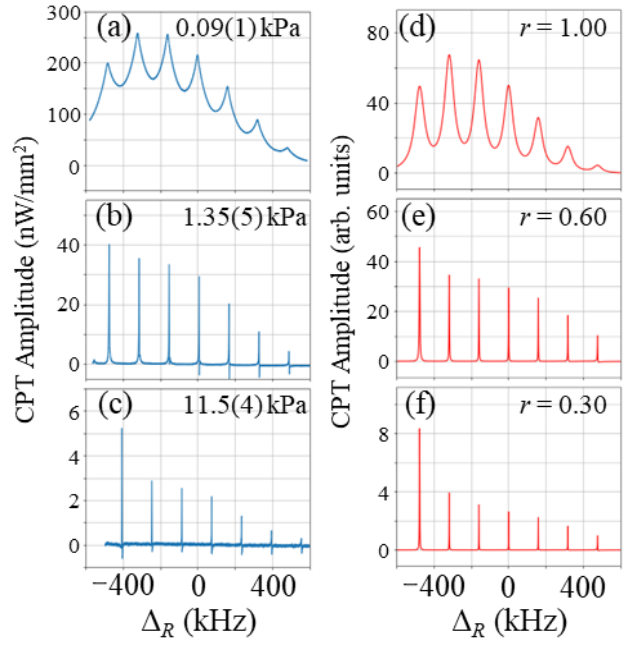

FIG. 5

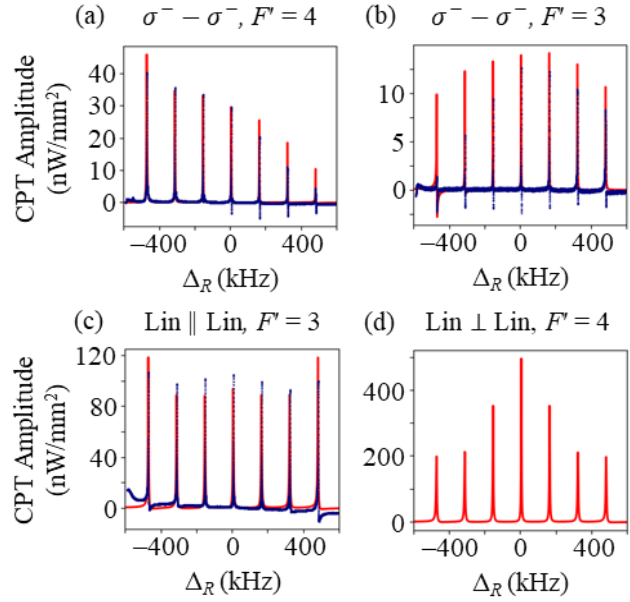

FIG. 6

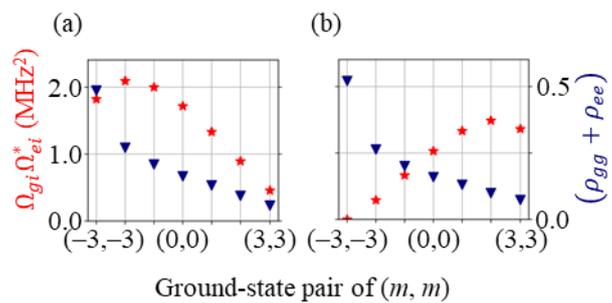

FIG. 7

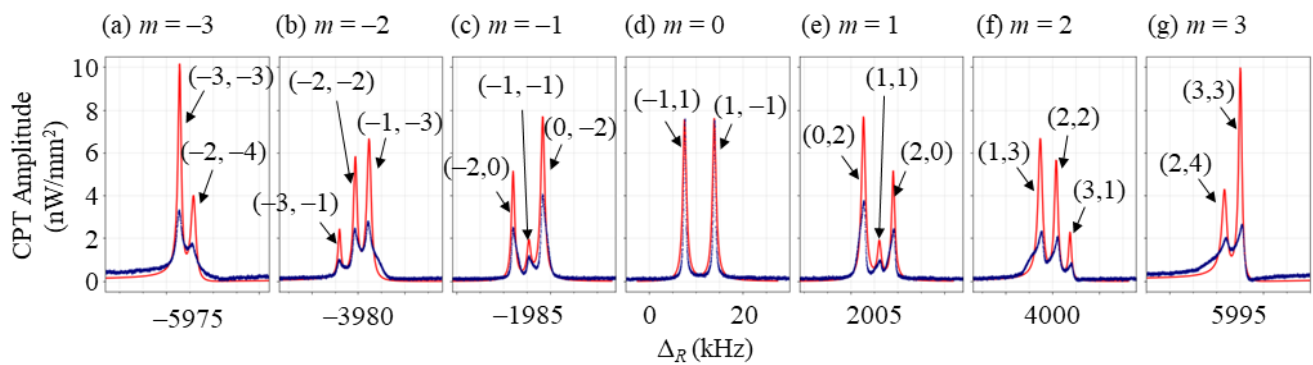

FIG. 8

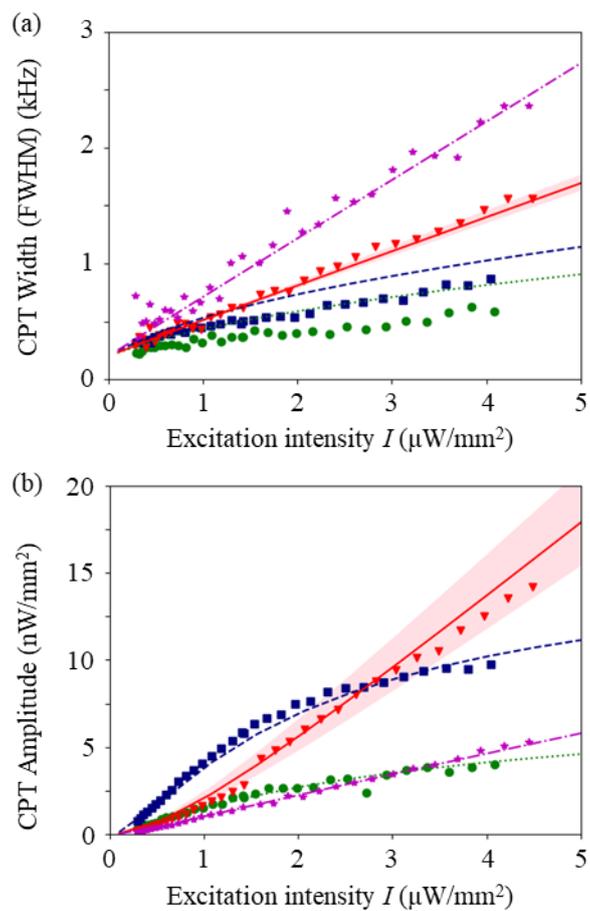

FIG. 9

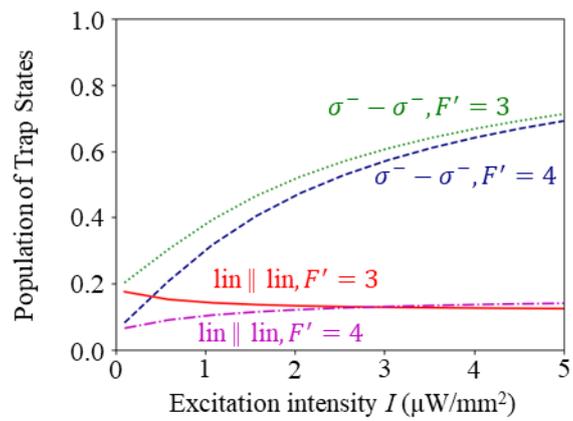